\documentclass[english]{sbrt}
\usepackage[english]{babel}
\usepackage[utf8]{inputenc}
\usepackage{graphicx}
\usepackage{amsmath,amssymb}
\usepackage{bm}
\usepackage{booktabs}
\usepackage{xcolor}
\usepackage{multirow}

\begin{document}

\title{FiPA-SR -- FiLM-Conditioned Perceptually Informed Audio Super-Resolution}

\author{Wallace Abreu and Luiz W. P. Biscainho
\thanks{Wallace Costa de Abreu, PEE/COPPE, UFRJ, Rio de Janeiro-RJ, e-mail: wallace.abreu@smt.ufrj.br; Luiz Wagner Pereira Biscainho, DEL/Poli \& PEE/COPPE, UFRJ, Rio de Janeiro-RJ, e-mail: wagner@smt.ufrj.br. This work was partially supported by the Brazilian Federal Agency for Support and Evaluation of Graduate Education, CAPES (001), the National Council for Scientific and Technological Development, CNPq (306395/2025-80), and the Carlos Chagas Filho Foundation for Research Support in the State of Rio de Janeiro, FAPERJ (E-26/204.092/2022).}%
}

\maketitle


\begin{abstract}
Audio bandwidth extension aims to reconstruct missing high-frequency content from bandlimited signals. This paper proposes FiPA-SR, a GAN-based perceptual architecture capable of handling different input bandwidths within a single model. Building upon the previous $\textrm{AEROMamba}_\textrm{P}$ framework, the proposed model incorporates FiLM layers to adapt the reconstruction process according to the respective bandwidth. Experiments on the MUSDB dataset show that FiPA-SR outperforms the state-of-the-art AudioSR model across 8, 20, and 32 kHz input sampling rates. Moreover, the proposed architecture uses approximately 3$\times$ less GPU memory and performs inference more than 60$\times$ faster than the diffusion-based baseline.
\end{abstract}

\section{Introduction}

Audio bandwidth extension, also known as audio super-resolution, refers to the reconstruction of missing high-frequency content of a bandlimited signal~\cite{Larsen04}. The suppression of high-frequency components often produces muffled or unnatural audio, motivating the development of techniques capable of reconstructing the upper spectrum from degraded inputs.

In several scenarios involving audio transmission and storage, bandwidth limitations are commonly introduced by communication systems~\cite{Valin00}, lossy compression~\cite{Brandenburg99} or media degradation~\cite{Copeland08}. For technologies that prioritize intelligibility~\cite{Coelho17}, such as telephony, moderate bandwidth limitations might be acceptable. However, in the case of music, high-fidelity systems must cover at least the human auditory range of 20 Hz to 20 kHz for tones~\cite{Bosi02}.

Classical bandwidth extension methods relied on signal processing techniques such as nonlinear devices with linear filtering~\cite{Larsen04}, source-filter modeling~\cite{Spanias06}, codebook mapping~\cite{Schmidt08}, and spectral-band replication~\cite{Ekstrand02}. More recently, deep learning approaches have become dominant in audio super-resolution, with the state-of-the-art solutions using generative approaches such as generative adversarial networks (GANs)~\cite{Abreu24, Abreu26}, diffusion models~\cite{Chen24,Moliner24} and latent bridge models~\cite{Li26}.

Although diffusion-based methods currently achieve state-of-the-art perceptual quality, their iterative sampling procedures~\cite{Ho2020DDPM} impose substantial computational costs during inference, making real-time or low-latency deployment challenging. When faster inference is needed, model distillation is a required step~\cite{Im25}.

In contrast, GAN-based approaches allow direct signal generation in a single forward pass, resulting in significantly lower inference complexity~\cite{Goodfellow14}. Motivated by this, recent work introduced $\textrm{AEROMamba}_\textrm{P}$~\cite{Abreu26}, a U-Net~\cite{Ronneberger15} based super-resolution architecture employing Mamba~\cite{Gu2024mamba}, a selective state space model designed for efficient sequence modeling, together with a differentiable Perceptual Audio Quality Measure (PAQM)~\cite{Beerends1992} loss for perceptually motivated training.

\begin{figure}[!h]
    \centering
    \includegraphics[width=0.67\linewidth]{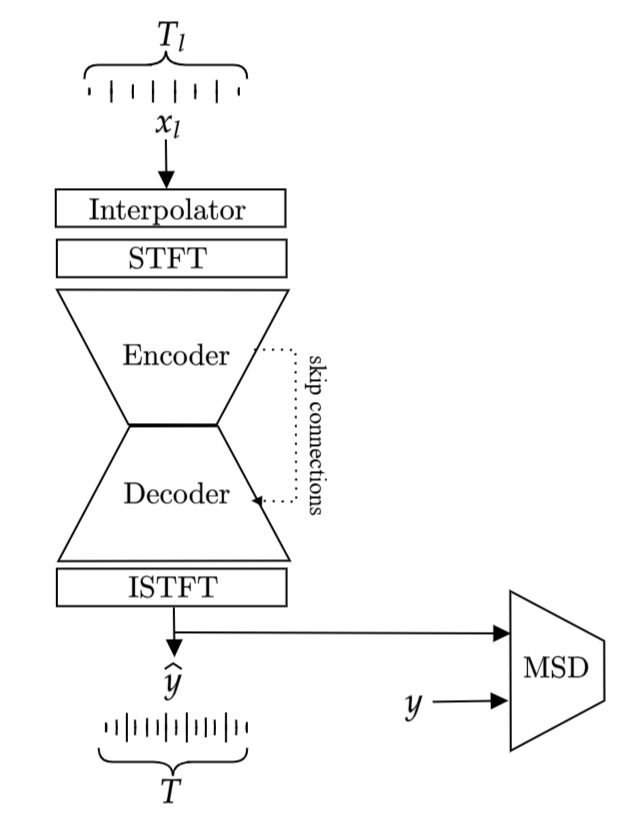}
    \caption{General FiPA-SR architecture, including the generator and discriminator.}
    \label{fig:FiPASR_diagram}
\end{figure}

However, unlike models such as AudioSR~\cite{Chen24}, which are capable of processing signals with multiple input sampling frequencies, $\textrm{AEROMamba}_\textrm{P}$ was designed for a single bandwidth configuration per training. To address this limitation, this work proposes FiPA-SR (FiLM-Conditioned Perceptual Audio Super-Resolution), illustrated in Figure~\ref{fig:FiPASR_diagram}, an extension of $\textrm{AEROMamba}_\textrm{P}$ that incorporates Feature-wise Linear Modulation (FiLM)~\cite{Perez18} layers, enabling a single model to reconstruct signals from multiple bandwidth configurations and enhance them to 44.1 kHz. 

The motivation for this modification arises from the fact that different cutoff frequencies preserve different amounts of spectral information, directly affecting the reconstruction task. Through FiLM conditioning, the network feature maps are adaptively modulated via affine transformations controlled by a conditional vector associated with the input sampling frequency, allowing the model to adjust its behavior according to the amount of spectral content available in the degraded signal.

The proposed model is evaluated in a bandwidth extension task considering three representative bandwidth configurations somewhat associated with communication systems: 4 kHz, 10 kHz, and 16 kHz. Experimental results demonstrate that FiPA-SR consistently outperforms both AudioSR and PA-SR (an ablation version of the model without FiLM) in the objective metrics Log-Spectral Distance (LSD)~\cite{Mandel22} and Virtual Speech Quality Objective Listener (ViSQOL)~\cite{Hines2015}. Moreover, the proposed architecture utilizes in inference one-third of AudioSR GPU, and produces outputs more than 60$\times$ faster. To ensure reproducibility and allow perceptual examination, a public repository and a webpage\footnote{https://fipa-sr.github.io/} accompany this work, containing the source code, model checkpoints, and audio examples.

\section{Method}

The FiPA-SR architecture is illustrated in Figure~\ref{fig:FiPASR_diagram}. Similar to its predecessor, $\textrm{AEROMamba}_\textrm{P}$~\cite{Abreu26}, FiPA-SR is a super-resolution GAN composed of a generator based on U-Net~\cite{Ronneberger15} and a multi-scale discriminator. Given a signal $\bm{y} \in \mathbb{R}^T$ and its downsampled version $\bm{x}_l \in \mathbb{R}^{T_l}$, where $T$ is the original signal length and $T_l$ is the downsampled signal length, FiPA-SR aims to generate $\hat{\bm{y}} \approx \bm{y}$, reconstructing the lost high-frequency content of $\bm{x}_l$. In practice, the input audio is transformed into a complex spectrogram using the STFT, treating real and imaginary components as different channels,  and processed by the network to estimate missing high-frequencies. The reconstructed spectrogram is then converted back to the time domain through an inverse STFT.

To handle different input sampling frequencies, a major modification introduced by FiPA-SR is the prior upsampling of the input signals to 44.1 kHz. Through this method, segments sampled at any frequency can generate spectrograms with the same size using identical parameters.  The low-resolution spectrograms can then be mapped by the model into a high-frequency version. This upsampling procedure employs an interpolation biquad filter of length 128 to minimize reconstruction artifacts.

A further enhancement to the model are FiLM layers, introduced at the entrance of the residual blocks, after layer normalization (LN). The FiLM layers apply an affine transformation to an input tensor $\bm{x}$ via a pair of scaling and shifting factors $\bm{\gamma} (\bm{c})$, $\bm{\beta} (\bm{c})$ conditioned by a vector $\bm{c}$. In this setting, $\bm{c}$ contains the normalized input sampling frequencies for each sample in the batch. Formally, for a batch of size $B$, a vector $\bm{c} = [c_1, c_2, \dots, c_B]^T$ is assigned,
where $c_i=\frac{f_s^{(i)}}{44100},\ i = 1, 2, \dots, B$. Since the accepted sampling frequencies are between 0 and 44100, then $0\leq c_i\leq1$.

\begin{figure}[!htp]
    \centering
    {\includegraphics[width=0.4\textwidth]{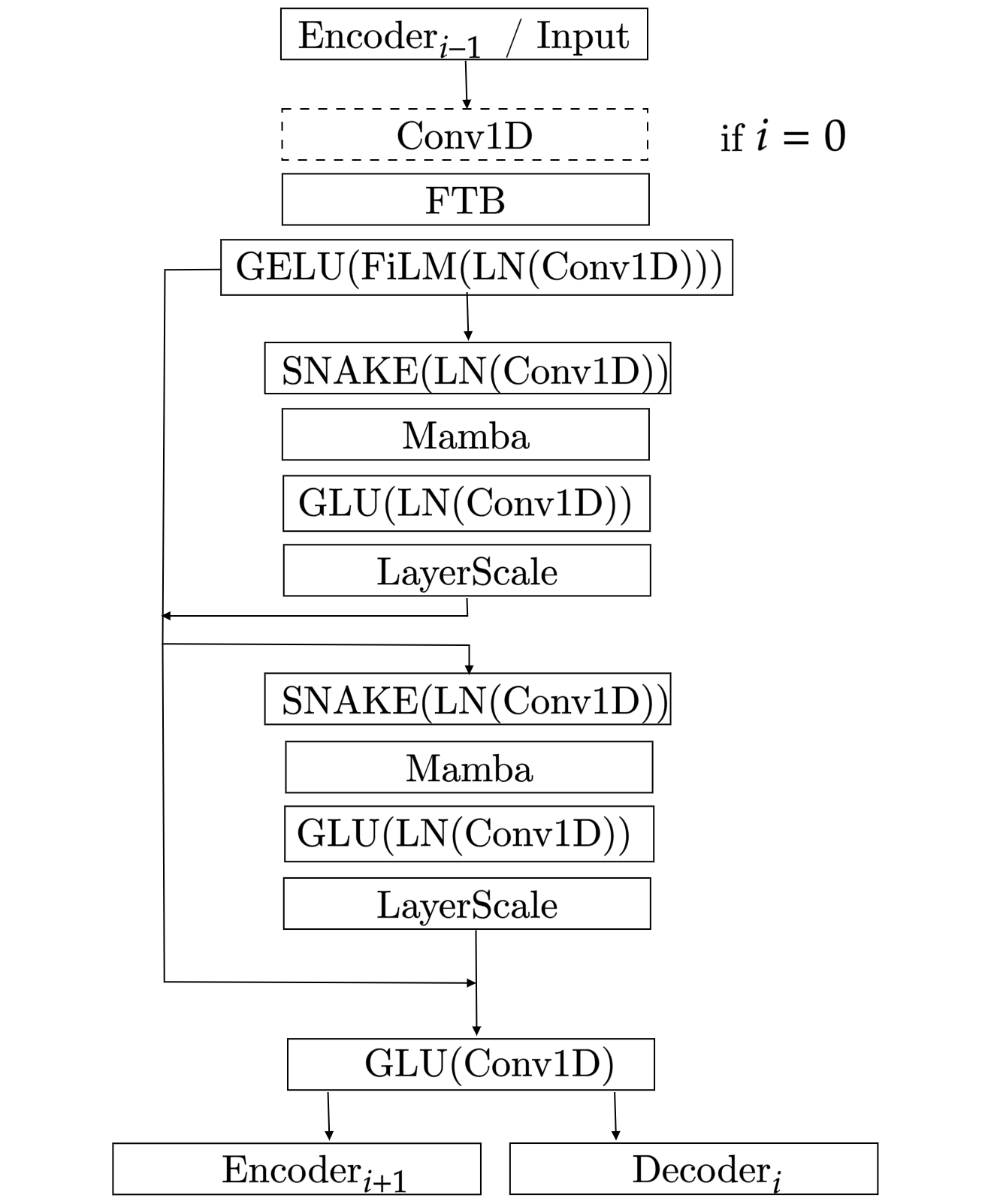}} 
    \caption{FiPA-SR encoder layer architecture.}
    \label{fig:FiPA-SR}
\end{figure}

In Figure~\ref{fig:FiPA-SR}, the complete encoder architecture is detailed. Its core comprises two identical residual blocks composed of Snake~\cite{Ziyin20} activation function, LN, Conv1D, Mamba and GLU. They are preceded by 1D convolutional layers followed by LN, the FiLM layer and a GELU. The decoder is symmetrical to the encoder and shares skip connections with the encoder layers. For adversarial training, FiPA-SR employs a MelGAN~\cite{Kumar19} multi-scale discriminator, an ensemble of three independent discriminators that operate directly on the waveform at different scales.

The proposed model, as in~\cite{Abreu26}, is trained using a composite generator loss defined as
\begin{equation}\label{eq:FiPASR_loss}
   L_\mathcal{G} = L_\text{adv} + L_\text{rec} + \lambda L_{\text{fmap}} - \gamma L_{\mathrm{PAQM}},
\end{equation}
where $L_\text{adv}$ denotes the adversarial loss, $L_\text{rec}$ is the spectral reconstruction loss, $L_\text{fmap}$ represents the feature matching loss, and $L_{\mathrm{PAQM}}$ corresponds to the PAQM score computed between the reconstructed and reference signals. The parameters $\lambda$ and $\gamma$ control the contribution of the feature matching and perceptual loss terms, respectively.
\section{Experiments}

\subsection{Dataset}
The MUSDB~\cite{musdb18-hq} dataset is composed of 150 full-length music tracks of different genres, along with their isolated drums, bass, vocals, and other stems, totaling approximately 10 hours in duration. The training folder contains 100 songs (around 7 hours), while the test folder contains 50 songs (around 3 hours), all of which are in WAV format, stereo mode, and sampled at 44.1 kHz. The number of tracks per genre is distributed as follows: 83 Pop/Rock, 24 Rock, 15 Singer/Songwriter, 14 Heavy Metal, 13 Pop, 10 Rap, 9 Electronic, 3 Country, 3 Jazz, and 2 Reggae tracks.

Here, as in previous works~\cite{Mandel22, Abreu26}, the test set was used in validation and test phases using independent metrics. Although using the same dataset is discouraged, this procedure was used to ensure compatibility with the baseline works.

\subsection{Overall Setup}
In the experiment, the outputs of different models were evaluated regarding their ability to enhance low-resolution signals with varied sampling frequencies. The respective input bandwidths were selected to reflect common communication channel limitations, namely: 4 kHz, typical of a telephone channel, and 16 kHz, approximating an FM broadcasting channel, along with an intermediate value of 10 kHz.

The assessed models were FiPA-SR, PA-SR, and the pretrained AudioSR model. Both FiPA-SR variants were jointly trained to handle the three considered bandwidth configurations within a single model, while AudioSR model was originally designed to support frequencies from 2 kHz up to 48 kHz. The difference between FiPA-SR and PA-SR lies in the removal of the FiLM layer in the latter, which is used as an ablation to evaluate the contribution of FiLM conditioning to the proposed architecture.

The pretrained AudioSR model served only as a supplementary baseline due to its extensive training of 7000 hours and the fact that it was exposed, in training, to the entire MUSDB dataset, including the test set, which the authors acknowledge that will benefit AudioSR in relation to the proposed models.

\subsection{Evaluation}
The models were evaluated using objective metrics: the Log-Spectral Distance (LSD)~\cite{Mandel22} and two distinct perceptual quality metrics, PAQM and the Virtual Speech Quality Objective Listener (ViSQOL)~\cite{Hines2015}.

PAQM was used as a validation metric to select the optimal model during a training run, with fixed seed and hyperparameters. The choice of PAQM as a validation metric was motivated by the availability of a vectorized implementation, which is much faster than ViSQOL, and also for its perceptual motivation, taking into account masking effects and other auditory modeling aspects~\cite{Beerends1992}. In addition, ViSQOL (in audio mode) was employed to assess the quality of the processed signals on a scale from 1 to 5, in relation to the test set.

The use of GPU by AudioSR and FiPA-SR was evaluated using the `nvidia-smi' on different audio tracks. Regarding inference speed, the time required to process 10-second segments was recorded, and the average duration for each method was calculated.

Average ViSQOL scores were tested for statistical significance and, except when explicitly stated, were considered to reject the null hypothesis of no significant difference with $p\mathrm{-value} < 0.05$.

\subsection{Training procedure}
FiPA-SR and PA-SR were trained for the bandwidth extension task considering the sampling frequencies of 8, 20, and 32 kHz. To enable a single model to handle multiple input sampling frequencies, the MUSDB dataset was resampled to each low-resolution frequency, and all tracks were jointly used during training. After frequency interpolation, the models employed a window size of $W=512$ and a hop length of $H=256$.

Both models were trained for 105 epochs using the default $\textrm{AEROMamba}_\mathrm{P}$ hyperparameters, as summarized in Table~\ref{table:train_parameters_wav}. Training was conducted on an NVIDIA RTX 3090 GPU with a batch size of 8. Finally, the model selection was based on the convergence of the PAQM score in the test set, with convergence defined as the situation where the score varies by less than 3\% over five consecutive epochs.
\begin{table}[t]
\centering
\renewcommand{\arraystretch}{1} 
\caption{Summary of the training parameters used in the experiments.}
\begin{tabular}{@{\hskip 2pt}lc@{\hskip 2pt}}
\toprule
\textbf{Parameter} & \textbf{Value} \\ \midrule
Window Size ($W$)     & 512 samples \\ 
Hop Length ($H$)      & 256 samples \\ 
Segment Length        & 4 s \\ 
Stride Length         & 4 s \\ 
Number of Epochs      & 105 \\ 
Batch Size            & 8 \\ 
Optimizer             & Adam~\cite{Kingma15} \\
Learning Rate         & $3 \times 10^{-4}$ \\ 
PAQM Loss Parameter ($\gamma$) & 1 \\
Feature Matching Parameter ($\lambda$) & 100 \\ 
\bottomrule
\end{tabular}
\label{table:train_parameters_wav}
\end{table}

\section{Results}

Results and computational performance metrics are summarized in Tables~\ref{tab:bandwidth_extension_results} and~\ref{table:model_complexity}, respectively. The proposed FiPA-SR achieved superior objective performance in fidelity compared to AudioSR and PA-SR, obtaining better LSD and ViSQOL scores across the evaluated bandwidth configurations. In addition, FiPA-SR demonstrated substantially lower computational complexity, requiring less than one-third of the GPU memory consumption and performing inference on 10-s audio segments more than 60 times faster than AudioSR. These advantages can be attributed both to the nearly 100$\times$ smaller number of parameters of FiPA-SR and to the inherently faster inference process of GAN-based models compared to diffusion-based sampling methods.
\begin{table*}[!htb]
\centering
\begin{tabular}{lcccccc}
\toprule
\multirow{2}{*}{Model} & \multicolumn{2}{c}{8 kHz} & \multicolumn{2}{c}{20 kHz} & \multicolumn{2}{c}{32 kHz} \\
\cmidrule(lr){2-3} \cmidrule(lr){4-5} \cmidrule(lr){6-7}
& ViSQOL $\uparrow$ & LSD $\downarrow$ 
& ViSQOL $\uparrow$ & LSD $\downarrow$
& ViSQOL $\uparrow$ & LSD $\downarrow$ \\
\midrule

Low-Resolution
& 1.64 & 2.17 
& 2.44 & 1.82 
& 4.10 & 1.64 \\

FiPA-SR 
& \textbf{2.82} & \textbf{1.24} 
& \textbf{3.53} & \textbf{1.04} 
& \textbf{4.41} & \textbf{0.68} \\

PA-SR
& 2.56 & 1.52 
& 3.12 & 1.19 
& 4.21 & 0.87 \\

AudioSR
& 2.72 & 1.69 
& 3.33 & 1.30 
& 3.85 & 1.06 \\
\bottomrule
\end{tabular}
\caption{Objective evaluation results for the bandwidth extension experiment for each input sampling frequency.}
\label{tab:bandwidth_extension_results}
\end{table*}
\begin{table}[ht]
\centering
\caption{GPU usage (VRAM), inference time for a 10-s segment, and total number of parameters for each model.}
\begin{tabular}{lccc}
\toprule
\textbf{Method} & \textbf{VRAM (MB)} & \textbf{Time (s)} & \textbf{Total Parameters} \\
\midrule
FiPA-SR & 3000 & 0.087 & 19,487,758 \\
AudioSR & 14396 & 5.663 & 1,285,395,637 \\
\bottomrule
\end{tabular}
\label{table:model_complexity}
\end{table}

\begin{figure}[!ht]
    \centering
    \includegraphics[width=1\linewidth]{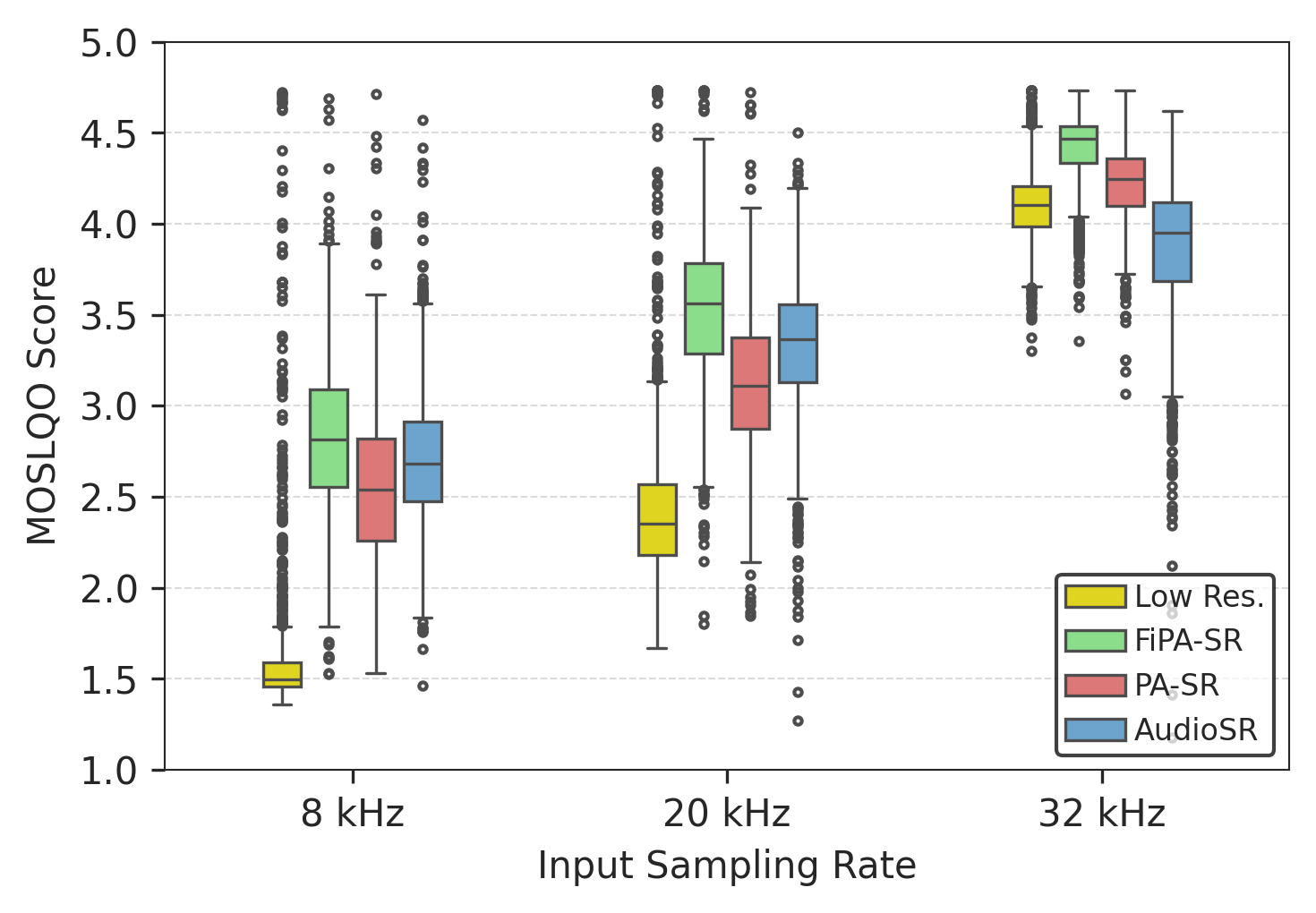}
    \caption{Distribution of ViSQOL scores of each low-resolution and model output for varying input sampling rates.}
    \label{fig:boxplot}
\end{figure}

The objective metrics also illustrate how FiLM layers are essential to the performance of FiPA-SR, especially at lower sampling rates. At 8 and 20 kHz, PA-SR yields lower ViSQOL scores than FiPA-SR and AudioSR, indicating that the removal of FiLM significantly degrades the model’s ability to generalize across multiple input bandwidth configurations. This shows that the model relies on the conditional vectors provided by FiLM to effectively create high-frequency content. In contrast, at 32 kHz, where the bandwidth extension task is less demanding due to the reduced amount of missing high-frequencies, PA-SR is able to surpass both AudioSR and baseline. A more detailed distribution of the ViSQOL results is presented in Figure~\ref{fig:boxplot}.

\begin{figure*}[!ht]
    \centering
    \includegraphics[width=0.9\linewidth]{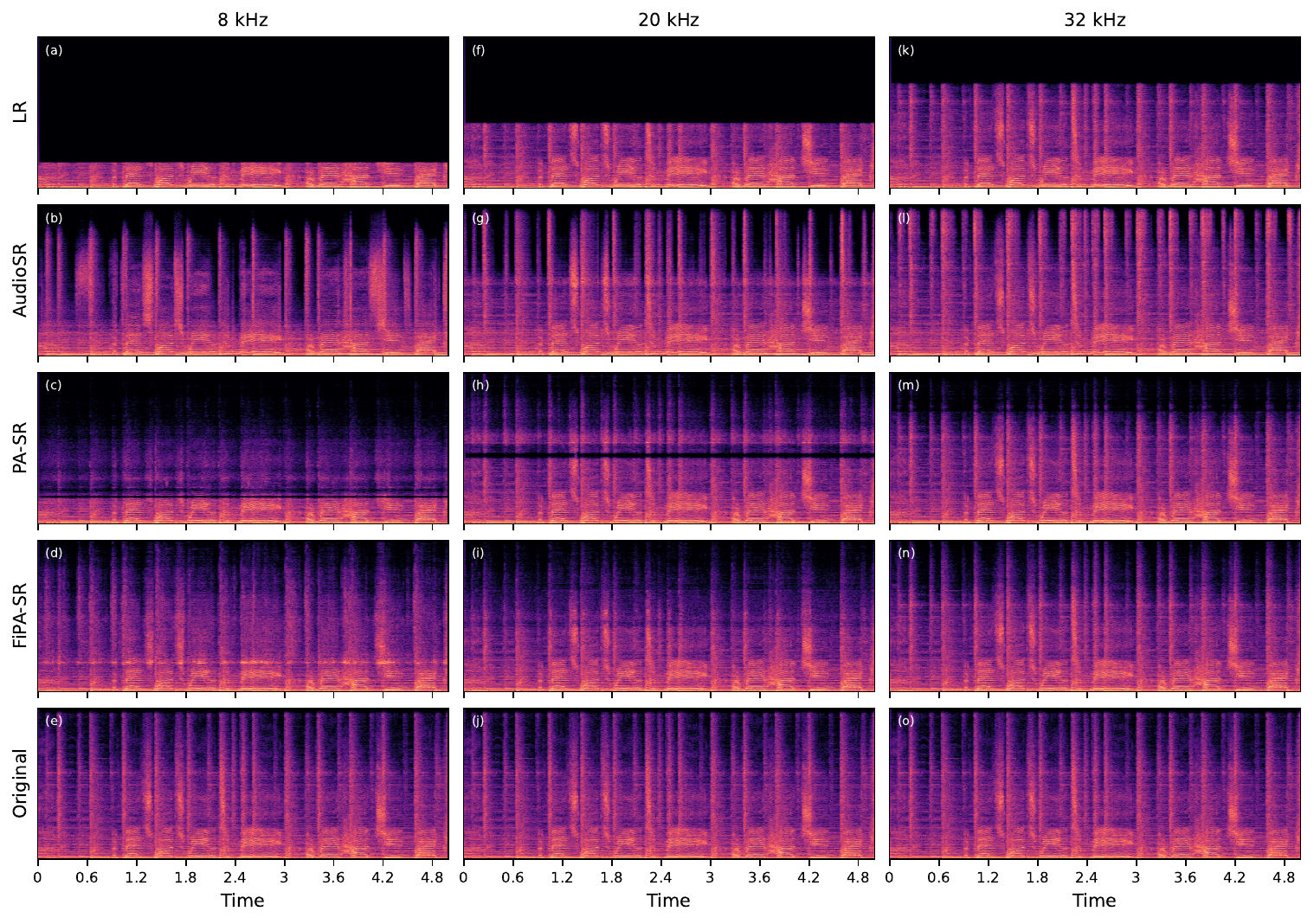}
    \caption{Spectrograms of a 5-second audio segment for different low-resolution signals and each model output in relation to an original 44.1 kHz signal.}
    \label{fig:spec_comparison}
\end{figure*}

The spectrogram visualizations shown in Figure~\ref{fig:spec_comparison} help illustrate the behavior of the models. The figure depicts a 5-second pop-rock segment in its original, low-resolution, and reconstructed versions for each evaluated model. AudioSR successfully extends the bandwidth for all input configurations, which is expected given its robustness as a state-of-the-art super-resolution model and its extensive training setup, including exposure to the complete MUSDB dataset. However, unlike approaches that aim for fidelity, AudioSR is primarily designed to generate perceptually plausible high-frequency content rather than faithfully reconstruct the original signal. Informal listening indicates that this behavior often leads to excessive percussive transients and noticeable alterations of timbre.

The difference between FiPA-SR and PA-SR further highlights the importance of FiLM conditioning in the proposed architecture. Beyond enabling FiPA-SR to effectively handle multiple input bandwidth configurations, the FiLM layer also completely removes spectral discontinuities near the cutoff regions. In contrast, PA-SR consistently exhibits visible border artifacts in the reconstructed high-frequency regions across all evaluated bandwidths. Informal listening indicates that the perceived audio quality is consistent with the ViSQOL results, with FiPA-SR producing natural-sounding reconstructions containing few perceptible artifacts for all the three configurations.

\section{Conclusions}
This paper introduced FiPA-SR, a GAN-based perceptual model for audio bandwidth extension capable of handling multiple input sampling rates. Building upon the previous $\textrm{AEROMamba}_\textrm{P}$ framework, the proposed model preserves computational efficiency while reconstructing high-frequency content across different bandwidth configurations. Experimental results demonstrated through objective metrics that FiPA-SR outperforms the state-of-the-art AudioSR model in all bandwidth settings, achieving from 0.10 to 0.56 higher MOSLQO in ViSQOL while using 3$\times$ less GPU memory and performing inference more than 60$\times$ faster.

The experiments also showed that a direct adaptation of AEROMamba to multiple bandwidth configurations, PA-SR, was not sufficient to achieve robust performance. In particular, the inclusion of FiLM layers proved essential, especially at lower sampling rates. The affine transformations controlled by the conditional vectors enabled the model to effectively generalize across different input bandwidths, while also reducing spectral discontinuities near the cutoff regions.

Future work will follow several main directions. First, the bandwidth extension experiments will be expanded to cover a broader range of sampling frequencies, from 4 kHz up to 48 kHz. In addition, subjective listening tests will be conducted to complement the objective evaluation and assess the perceptual quality of the reconstructed audio. Second, the applicability of the proposed model will be investigated in related audio processing tasks, such as encoded audio enhancement, restoration of archival recordings, and automatic equalization. Finally, from an explainability perspective, future studies will focus on understanding how different audio characteristics, such as percussive-dominant, harmonic-dominant, transient-rich, and isolated-instrument content, influence the behavior and performance of the model.




\begin{thebibliography}{99}
\bibitem{Larsen04}
Larsen, E., \& Aarts, R.
Audio Bandwidth Extension.
(John Wiley \& Sons, 2004)

\bibitem{Valin00}
Valin, J., \& Lefebvre, R.
Bandwidth Extension of Narrowband Speech for Low Bit-rate Wideband Coding.
{\em Proceedings of the IEEE Speech Coding Workshop}.
pp. 130--132 (2000)

\bibitem{Brandenburg99}
Brandenburg, K.
MP3 and AAC Explained.
{\em Proceedings of the AES 17th International Conference on High-Quality Audio Coding}.
pp. 99--110 (1999)

\bibitem{Copeland08}
Copeland, P.
Manual of Analogue Sound Restoration Techniques.
(British Library, 2008)

\bibitem{Coelho17}
Biscainho, L., \& Nunes, L.
Automatic Evaluation of Acoustically Degraded Full-band Speech.
{\em Signals and Images: Advances and Results in Speech, Estimation, Compression, Recognition, Filtering, and Processing}.
pp. 181--208 (2017)

\bibitem{Bosi02}
Bosi, M., \& Goldberg, R.
Introduction to Digital Audio Coding and Standards.
(Kluwer Academic Publishers, 2002)

\bibitem{Spanias06}
Berisha, V., \& Spanias, A.
Bandwidth Extension of Audio Based on Partial Loudness Criteria.
{\em Proceedings of the 8th IEEE Workshop on Multimedia Signal Processing}.
pp. 146--149 (2006)

\bibitem{Schmidt08}
Iser, B., \& Schmidt, G.
Bandwidth Extension of Telephony Speech.
(Springer, 2008)

\bibitem{Ekstrand02}
Ekstrand, P.
Bandwidth Extension of Audio Signals by Spectral Band Replication.
{\em Proceedings of the 1st IEEE Benelux Workshop on Model Based Processing and Coding of Audio}.
pp. 73--79 (2002)

\bibitem{Mandel22}
Mandel, M., Tal, O., \& Adi, Y.
AERO: Audio Super Resolution in the Spectral Domain.
{\em Proceedings of the IEEE International Conference on Acoustics, Speech and Signal Processing}.
(2023)

\bibitem{Abreu24}
Abreu, W., \& Biscainho, L.
AEROMamba: An Efficient Architecture for Audio Super-Resolution Using Generative Adversarial Networks and State Space Models.
{\em Proceedings of the 1st Latin American Music Information Retrieval Workshop}.
pp. 74--79 (2024)

\bibitem{Abreu26}
Abreu, W., Miranda, B. V., \& Biscainho, L. W. P.
Efficient Audio Enhancement With a Differentiable Psychoacoustic Loss.
{\em Journal of the Audio Engineering Society}.
\textbf{74}, 430--444 (2026)

\bibitem{Chen24}
Liu, H., Chen, K., Tian, Q., Wang, W., \& Plumbley, M.
AudioSR: Versatile Audio Super-resolution at Scale.
{\em Proceedings of the IEEE International Conference on Acoustics, Speech and Signal Processing}.
pp. 1076--1080 (2024)

\bibitem{Moliner24}
Moliner, E., Elvander, F., \& Välimäki, V.
Blind Audio Bandwidth Extension: A Diffusion-based Zero-shot Approach.
{\em IEEE/ACM Transactions on Audio, Speech, and Language Processing}.
\textbf{32}, 5092--5105 (2024)

\bibitem{Li26}
Li, C., Chen, Z., Wang, L., \& Zhu, J.
Audio Super-Resolution with Latent Bridge Models.
{\em Proceedings of the Thirty-ninth Annual Conference on Neural Information Processing Systems}.
(2026)

\bibitem{Ho2020DDPM}
Ho, J., Jain, A., \& Abbeel, P.
Denoising Diffusion Probabilistic Models.
{\em Advances in Neural Information Processing Systems}.
\textbf{33} (2020)

\bibitem{Im25}
Im, J., \& Nam, J.
FlashSR: One-step Versatile Audio Super-resolution via Diffusion Distillation.
{\em ICASSP 2025 -- IEEE International Conference on Acoustics, Speech and Signal Processing}.
pp. 1--5 (2025)

\bibitem{Goodfellow14}
Goodfellow, I., et al.
Generative Adversarial Nets.
{\em Proceedings of the 28th Annual Conference on Neural Information Processing Systems}.
pp. 2672--2680 (2014)

\bibitem{Ronneberger15}
Ronneberger, O., Fischer, P., \& Brox, T.
U-Net: Convolutional Networks for Biomedical Image Segmentation.
{\em Proceedings of the 18th International Conference on Medical Image Computing and Computer-Assisted Intervention}.
pp. 234--241 (2015)

\bibitem{Gu2024mamba}
Gu, A., \& Dao, T.
Mamba: Linear-Time Sequence Modeling with Selective State Spaces.
Available at: https://arxiv.org/abs/2312.00752 (2024)

\bibitem{Perez18}
Perez, E., Strub, F., Vries, H., Dumoulin, V., \& Courville, A.
FiLM: Visual Reasoning with a General Conditioning Layer.
{\em Proceedings of the Thirty-Second AAAI Conference on Artificial Intelligence}.
(2018)

\bibitem{Hines2015}
Hines, A., Skoglund, J., Kokaram, A., \& Harte, N.
ViSQOL: An Objective Speech Quality Model.
{\em EURASIP Journal on Audio, Speech, and Music Processing}.
pp. 1--18 (2015)

\bibitem{Ziyin20}
Ziyin, L., Hartwig, T., \& Ueda, M.
Neural Networks Fail to Learn Periodic Functions and How to Fix It.
{\em Proceedings of the 34th International Conference on Neural Information Processing Systems}.
pp. 1583--1594 (2020)

\bibitem{Kumar19}
Kumar, K., et al.
MelGAN: Generative Adversarial Networks for Conditional Waveform Synthesis.
{\em Proceedings of the 33rd International Conference on Neural Information Processing Systems}.
pp. 14910--14921 (2019)

\bibitem{musdb18-hq}
Rafii, Z., Liutkus, A., Stöter, F.-R., Mimilakis, S. I., \& Bittner, R.
MUSDB18-HQ -- An Uncompressed Version of MUSDB18.
Zenodo.
https://doi.org/10.5281/zenodo.3338373
(2019)

\bibitem{Beerends1992}
Beerends, J., \& Stemerdink, J.
A Perceptual Audio Quality Measure Based on a Psychoacoustic Sound Representation.
{\em Journal of the Audio Engineering Society}.
\textbf{40}, 963--978 (1992)

\bibitem{Kingma15}
Kingma, D., \& Ba, J.
Adam: A Method for Stochastic Optimization.
{\em Proceedings of the 3rd International Conference on Learning Representations}.
(2015)
\end{thebibliography}
\end{document}